\documentclass[aps,prb,twocolumn,floatfix,showpacs,superscriptaddress]{revtex4}

\usepackage{graphicx}
\usepackage{bbm}
\usepackage{amsmath,amssymb}

\newcommand{\nn} {\nonumber}
\newcommand{\be}{\begin{equation}}
\newcommand{\ee}{\end{equation}}
\newcommand{\bea}{\begin{eqnarray}}
\newcommand{\eea}{\end{eqnarray}}
\renewcommand{\vr} {{\bf r}}
\newcommand{\vs} {{\bf s}}

\begin{document}
\title{Orbital-free energy functional for electrons in two dimensions}
\author{S. Pittalis}
\email[Electronic address:\;]{pittaliss@missouri.edu}
\affiliation{Institut f{\"u}r Theoretische Physik,
Freie Universit{\"a}t Berlin, Arnimallee 14, D-14195 Berlin, Germany}
\affiliation{European Theoretical Spectroscopy Facility (ETSF)}
\author{E. R{\"a}s{\"a}nen}
\email[Electronic address:\;]{erasanen@jyu.fi}
\affiliation{Nanoscience Center, Department of Physics, University of
  Jyv\"askyl\"a, FI-40014 Jyv\"askyl\"a, Finland}
\affiliation{Institut f{\"u}r Theoretische Physik,
Freie Universit{\"a}t Berlin, Arnimallee 14, D-14195 Berlin, Germany}
\affiliation{European Theoretical Spectroscopy Facility (ETSF)}
\date{\today}

\begin{abstract}
We derive a non-empirical, orbital-free density functional for
the total energy of interacting electrons in two dimensions. 
The functional consists of a local formula for the interaction energy, 
where we follow the lines introduced by Parr for three-dimensional 
systems [R. G. Parr, J. Phys. Chem.  {\bf 92}, 3060 (1988)], 
and the Thomas-Fermi approximation for the kinetic energy.
The freedom from orbitals and from the Hartree integral 
makes the proposed approximation numerically highly efficient. 
The total energies obtained for confined two-dimensional systems are 
in a good agreement with the standard local-density approximation within
density-functional theory, and considerably more accurate than the
Thomas-Fermi approximation.
\end{abstract}

 
\maketitle

\section{Introduction}

Two-dimensional (2D) electronic systems have attracted
vast interest since the beginning of semiconductor 
technology. Important examples are quantum-Hall systems
and different types of quantum-dot (QD) devices.~\cite{jacak}
Technological development has also increased 
the need for computational methods capable to deal with the 
many-electron problem in reduced dimensions. 
Among the available methods is the well-known 
local-density approximation (LDA) within
the celebrated density-functional theory~\cite{dft} (DFT).
The 2D-LDA consists of the exchange functional derived
for the homogeneous 2D electron gas,~\cite{rajagopal} and the 
corresponding correlation functional constructed using quantum Monte Carlo 
methods.~\cite{tanatar,attaccalite} At present, DFT with
the 2D-LDA, and especially their spin-dependent 
(and current-dependent) extensions, are among the
standard methods in the electronic-structure 
calculations of semiconductor QD's.~\cite{reimann_rmp}
Further developments of 2D density functionals have begun
very recently for both exchange~\cite{x2D} and 
correlation.~\cite{c2D1,c2D2}

Although the LDA, for example, is an {\em explicit} density functional, so
that the total density is the sole input variable (instead of the electronic
orbitals), the standard Kohn-Sham (KS) scheme in DFT still requires 
the computation of the KS orbitals for the single-particle kinetic
energy. This sets limitations to the number of electrons 
that can be treated numerically. The so-called orbital-free 
DFT~\cite{Wang,Lign,Carter} scales 
better in this respect, but within this approach 
the construction of an accurate
energy functionals (in particular for 3D systems) has 
resulted to be a complicated task. 
The ``traditional'' Thomas-Fermi (TF) approximation may 
serve as an important example of an orbital-free functional.
The TF approach has been put on a 
mathematically rigorous basis,~\cite{lieb_rmp} and also analyzed in 2D 
in detail by Lieb {\em et al}.~\cite{lieb} 
Furthermore, the TF theory 
has been successfully applied in the electronic-structure 
calculations of, e.g., quantum-Hall systems, where the importance of 
{\em e-e} interactions has been addressed.~\cite{afif} 
The TF energy functional has, however, the obvious deficiency
to treat the {\em e-e} interaction only classically
(i.e., only Hartree energy in included). Therefore, in the regime of small number of particles 
and/or low densities (strong interactions) the performance of the TF
method is highly questionable due to the lack of quantum mechanical effects (exchange and
correlation).

In this paper we aim at 
bridging the gap between the numerical
efficiency of TF method, and the accuracy 
of standard KS-DFT, for 
electronic-structure calculations in 2D. To this end, we present an 
explicit density
functional for the total energy which accounts for the classical and, 
to some extent, also for the quantum mechanical contribution 
to the interaction energy. 
In the derivation for the interaction energy we follow the general lines
already employed in the 3D case by Parr.~\cite{RGP}
In particular, we apply a Gaussian approximation
of the second-order density matrix~\cite{c2D1} and make use of the properties
of the interaction energy under a scaling transformation. 
Combining the resulting formula with the TF approximation for the kinetic
energy leads to an explicit density functional 
for the total energy. Applications to
2D QD's and rectangular quantum slabs (QS's) up to 
hundreds of electrons show a significant 
improvement over the TF energies when compared with the LDA results
in a wide range of the {\em e-e} interaction strength.

\section{Derivation of the approximation}

Our aim is to obtain a good estimation of the total
energy of a 2D system with a large number of electrons using 
a simple and computationally convenient formula.

First, let us consider the the {\em e-e} interaction energy. This can be expressed 
in terms of the spinless second-order density matrix as
\be\label{W}
W= \int{d \vr_1} \int{d \vr_2}~\frac{\rho_2(\vr_1,\vr_2)}{|\vr_1-\vr_2|},
\ee
where
\bea
\rho_2(\vr_1,\vr_2) &=& \frac{N(N-1) }{2}
\sum_{\sigma_1,\sigma_2} \int{d 3}...\int dN \nn \\
&\times& |\Psi(\vr_1 \sigma_1,\vr_2 \sigma_2,3, ...,N)|^2.
\eea
Here, $\Psi(1,2,...,N)$ stands for the ground-state
many-body wave function and $\int dN$ denotes the spatial integration
and spin summation over the $N$th spatial spin coordinate $(\vr_N
\sigma_N)$.
Hartree atomic units are used throughout the paper 
unless stated otherwise. 
The above definition implies the normalization
\be\label{N}
\frac{N(N-1)}{2} = \int{d \vr_1} \int{d \vr_2} ~ \rho_2(\vr_1,\vr_2),
\ee
and $\rho_2$ can be interpreted as the distribution density of the electronic pairs.

Next, we will specialize all the expressions to the 2D case and 
derive a local-density approximation for the interaction energy $W$
defined in Eq. (\ref{W}). 
In the average, $\vr=(\vr_1+\vr_2)/2$, and relative, $\vs=\vr_1-\vr_2$, 
coordinates, Eq. (\ref{W}) can be rewritten as
\be\label{W2}
W = 2\pi \int{d \vr} \int{d s}~ \rho_2(\vr,s),
\ee
where we have introduced the cylindrical average of $\rho_2$, which is
defined as
\be
\rho_2(\vr,s) = \frac{1}{2 \pi} \int_{0}^{2 \pi}{d \phi_s}~\rho_2\left(\vr+\frac{\vs}{2},\vr-\frac{\vs}{2}\right).
\ee
We assume a Gaussian approximation to be valid for the 
cylindrical average of the pair-density distribution function,
\be\label{ga}
\rho_2(\vr,s) \approx \rho_2(\vr,\vr) \exp\left[-\frac{s^2}{\beta_2(\vr)}\right],
\ee
where we have introduced $\beta_2(\vr)$ as a quantity to be determined below.
Substituting Eq. (\ref{ga}) in Eq. (\ref{W2}), and integrating over the
relative coordinate, we obtain
\be\label{W3}
W = \pi^{3/2} \int d \vr \,\rho_2(\vr,\vr)\,\beta_2^{1/2}(\vr).
\ee
Similarly, substituting Eq. (\ref{ga}) in Eq. (\ref{N}) we obtain
\be\label{N3}
N(N-1) = 2 \pi  \int d \vr\, \rho_2(\vr,\vr)\,\beta_2(\vr).
\ee

An additional and crucial assumption is introduced by imposing 
the integrands of Eqs. (\ref{W3}) and (\ref{N3}), respectively,
to be dependent on the space variable through the particle density. 
Thus, we may write
\be\label{d1}
\rho_2(\vr,\vr) = \rho_2(\rho(\vr)),
\ee
and
\be\label{d2}
\beta_2(\vr) = \beta_2(\rho(\vr)).
\ee
It is possible to work out the dependencies on the particle densities
of the above quantities by a dimensional argument.
Under uniform scaling of the coordinates, $\vr \rightarrow \lambda \vr$ (with $0 < \lambda
< \infty$), and of the norm-preserving many-body wavefunction
\be\label{scwav}
\Psi_\lambda(\vr_1,...,\vr_N) = \lambda^{N} \Psi  (\lambda \vr_1,..., \lambda \vr_N).
\ee
As a consequence, the other quantities of interest scale as
\be
\rho_{2,\lambda}(\vr_1, \vr_2) = \lambda^4 \rho_2(\lambda \vr_1, \lambda \vr_2),
\ee
\be\label{scden}
\rho_{\lambda}(\vr) = \lambda^2 \rho(\lambda \vr),
\ee
and
\be
W[\Psi_\lambda] = \lambda W[\Psi].
\ee
By using the assumptions in Eqs. (\ref{d1}) and (\ref{d2}) together with 
the scaling properties listed above, and by a dimensional argument, we 
arrive at the following expressions
for the integrands in Eqs.~(\ref{W3}) and (\ref{N3}), respectively:
\be\label{I}
\rho_2(\vr,\vr) \,\beta_2^{1/3}(\vr) = C_1\, \rho^{3/2}(\vr),
\ee
and
\be\label{II}
\rho_2(\vr,\vr)\, \beta_2(\vr) = C_2\,  \rho(\vr),
\ee
where $C_1$, and $C_2$ are constants. Eqs. (\ref{I}) and (\ref{II}) imply
\be
\beta_2(\vr) = \frac{C^2_2}{C^2_1} \,\rho^{-1}(\vr) = A \,\rho^{-1}(\vr),
\ee
and
\be
\rho_2(\vr,\vr) = \frac{C_1^2}{C_2} \,\rho^2(\vr) = B \,\rho^2(\vr).
\ee
An estimation for the latter factor $B$ can be obtained by
considering the Hartree-Fock (HF) case, for which
\be
\label{coeff}
\rho_{2,{\rm HF}}(\vr,\vr) = \frac{1}{4}\, \rho^2_{\rm HF}(\vr).
\ee
Hence, $B=1/4$.
The other factor $A$ can be determined by imposing the
normalization condition in Eq. (\ref{N3}),
\be
A = \frac{2(N-1)}{\pi}.
\ee
Now we have all the information to give an explicit expression 
for the interaction energy, which results to be
\be\label{int}
W[\rho] = \frac{\pi}{2} \sqrt{\frac{N-1}{2}} \int d \vr \,\rho^{3/2}(\vr). 
\ee
We emphasize that the expression gives $W=0$ for $N=1$, 
while this is not recovered by the LDA and the TF approximations.
Of course, for a large electron number ($N \gg 1$) 
the above expression can be simplified as $N - 1 \approx N$.

An interesting feature in Eq. (\ref{int}) is the fact that it 
allows us to approximate the total {\em e-e} interaction 
in a very simple fashion, which
is computationally appealing for systems with a 
large number of electrons.
However, some caution is in order: 
In the derivation above, we have introduced the 
assumption in Eq. (\ref{d1})
and then invoked the HF case in determining the coefficient $B$ in
Eq.~(\ref{coeff}). Alternatively, one may refer to the exact exchange 
approximation within DFT. In any case, Eq.~(\ref{coeff}) 
is valid for a pair density 
matrix coming from a wavefunction of the form of a {\em single} Slater 
determinant. As a consequence, the resulting approximation may result
to be biased toward the Hartree plus exchange energy. Nevertheless, we
make this choice for methodological simplicity. Moreover, as shown below, the
resulting approximation allows to deal with strongly interacting
systems to a very good extent. Alternative choice for $B$ could be
made by considering correlated pair densities of reference systems,
such as the homogeneous 2D electron gas,~\cite{gorigiorgi1}
or by using a coupling-constant average wich allows to account 
for the correlation contribution to the kinetic energy.~\cite{gorigiorgi2}

We also point out that Eq.~(\ref{int}) has the disadvantage to define 
a functional that is not size-consistent. In fact, because of its 
nonlinear dependence on the particle number $N$, even in the case
that the exact kinetic energy would be known, the total energy of 
two non-interacting fragments is not equal to the sum of 
the two fragment energies calculated separately.

Now, as a simple approximation for the many-body
kinetic energy we propose the TF expression
\be
T_{\rm TF}[\rho] = \frac{\pi}{2} \int d \vr\, \rho^2(\vr).
\label{tf}
\ee 
First of all, it is reassuring to see that the TF kinetic energy scales 
as the exact one. In fact, from Eqs. (\ref{scwav}) and (\ref{scden})
it is straightforward to verify that
$T[\Psi_\lambda] = \lambda^2 T[\Psi]$, and 
$T_{\rm TF}[\rho_\lambda] = \lambda^2 T_{\rm TF}[\rho]$.
Moreover, in 2D the TF kinetic 
energy functional is particularly attractive, since 
its gradient corrections vanish to all 
orders,~\cite{Brack,march,salasnich,berkane} 
whereas in 3D the first $\hbar^2$ order correction is the well-known 
von Weizs\"acker correction.~\cite{weizsacker} 
Besides, for the 2D Fermi gas in a harmonic trap the 
TF kinetic energy yields the exact non-interacting kinetic energy 
when the exact density is used as the input.~\cite{Brack} 
But in interacting systems, even in the best case the present 
approximation misses the correlation
contribution to the kinetic energy. As mentioned above, it could 
be possible to account for this contribution by
introducing a coupling-constant average, which is, however, 
beyond the scope of this work.

Combination of Eq.~(\ref{int}) with Eq.~(\ref{tf}) yields an 
orbital-free density functional for the {\em total} energy,
\bea\label{etot}
E[\rho] & = & T_{\rm TF}[\rho] + \frac{\pi}{2} \sqrt{\frac{N-1}{2}} \int d \vr \,\rho^{3/2}(\vr)  \nn \\
& + & \int d \vr\, \rho(\vr)\,v_{\rm  ext}(\vr).
\eea

We remind that the standard TF approximation for the total
energy is given by
\be\label{tfenergy}
E_{\rm TF}[\rho]=T_{\rm TF}[\rho]+E_H[\rho]  +  \int d \vr\, \rho(\vr)\,v_{\rm  ext}(\vr),
\ee
where the Hartree energy is defined by
\be\label{hartree}
E_H[\rho]=\frac{1}{2}\int d\vr \int d\vr'\,\frac{\rho(\vr)\rho(\vr')}{|\vr-\vr'|}.
\ee
In the LDA the total energy has an expression
\bea
E_{\rm LDA}[\rho] &=& T_{\rm KS}[\rho]+ E_H[\rho]+E^{\rm LDA}_x[\rho]+E^{\rm LDA}_c[\rho]  \nn \\
& + & \int d \vr\, \rho(\vr)\,v_{\rm  ext}(\vr),
\label{lda}
\eea
where the KS kinetic energy is calculated from the KS {\em orbitals}.
Therefore, the LDA expression is an {\em implicit} density functional
in contrast with the explicit density functionals in 
Eqs.~(\ref{etot}) and (\ref{tfenergy}).

\section{Numerical procedure}

We test our total-energy functional
given in Eq.~(\ref{etot}) 
on parabolic (harmonic) QD's and rectangular QS's, respectively.
The QD is defined by a harmonic external confining potential 
$v_{\rm ext}(r)=\omega^2 r^2 /2$ on the {\em xy} plane, 
where $\omega$ is the confinement
strength. The average electron density
in the QD can be approximated by a density parameter 
$r_s=N^{-1/6}\omega^{-2/3}$ (Ref.~\onlinecite{koskinen}).
The parameter corresponds to the 
average radius of an electron in a QD with an average number 
density $n_0=1/(\pi r_s^2)$. In the case of the QS, the confining
potential is a 2D rectangular quantum well with steep (hard-wall) 
boundaries,~\cite{recta1,recta2}
and the density parameter can be determined from 
$r_s=\sqrt{A/(\pi N)}$, where $A$ is the area of the QS. 

For both test systems, in the parameter ranges considered here
for $N$, $\omega$, and $A$, the LDA is known to yield very good 
total energies in comparison with exact or semi-exact many-electron
methods, e.g., quantum Monte Carlo calculations.~\cite{lda,exx_lda,recta1}
Therefore we use the LDA energies as the reference data
in this work. We compute the LDA energies in the standard 
KS scheme by applying {\tt octopus} real-space DFT code.~\cite{octopus} 
For the LDA correlation [last term in Eq.~(\ref{lda})]
we use the parametrization of 
Attaccalite {\em et al.}~\cite{attaccalite} 

We apply our functional in Eq.~(\ref{etot}), as well
as the TF expression in Eq.~(\ref{tfenergy}), by using
the self-consistent LDA density as the input density in
a one-shot calculation. In this way, all the functionals 
are evaluated with the same particle densities, and the 
LDA densities may be considered as reasonable estimations 
of the exact ones. The possibility for a self-consistent 
application of our functional, and its practical relevance,
are subjects of future investigation.

\section{Results}\label{totresults}

\begin{figure}
\includegraphics[width=0.8\columnwidth]{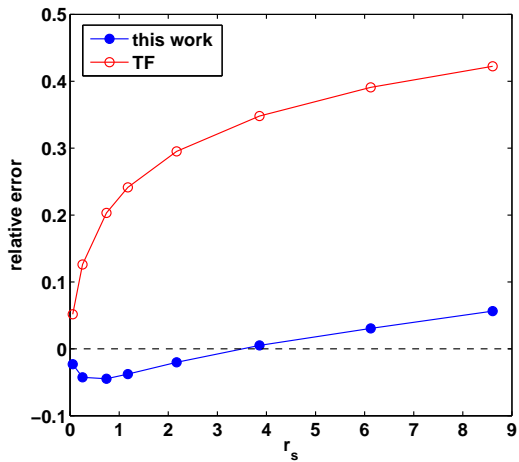}
\caption{(color online)
Relative error in the total energy
calculated with our functional (filled circles)
and with the Thomas-Fermi approximation (open circles)
for a six-electron parabolic quantum dot as a 
function of the density parameter $r_s$.
}
\label{fig1}
\end{figure}

First we consider the total energies
of a spin-unpolarized quantum dot with a fixed
number of electrons, $N=6$, as a function of the density 
parameter $r_s$.
Figure~\ref{fig1}
shows the relative error of our functional 
(filled circles) and the TF approximation (open circles)
in the total energy against the reference LDA result, i.e.,
$|E_{\rm LDA}-E|/E_{\rm LDA}$. We emphasize that for this
system the relative error of the reference LDA is below 
$0.003$ with respect to quantum Monte Carlo 
calculations.~\cite{lda} The values used for $r_s$ correspond to
a wide range of the interaction strength, covering well
the typical values ($r_s\sim 1 \ldots 5$) used when modeling 
QD's within the effective-mass approximation at
the interface of GaAs and AlGaAs.~\cite{jacak,reimann_rmp,kouwenhoven}

Overall, we find an excellent agreement in the total energies
between the LDA and our functional. The relative error remains
below $\sim 5\,\% $ through the full range of $r_s$. On the other
hand, the TF approximation is accurate only close to the 
noninteracting limit ($r_s\rightarrow 0$), whereas in general the
TF error is dozens of percent. The overestimation of the total energy
in the TF approximation is plausible due to the lack of exchange
and correlation energies which are both always negative.

\begin{figure}
\includegraphics[width=0.9\columnwidth]{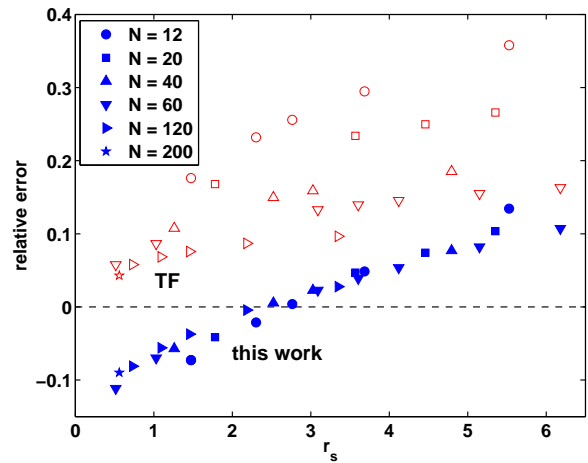}
\caption{(color online) Relative error in the total energy
calculated with our functional (filled symbols)
and with the Thomas-Fermi approximation (open symbols)
for rectangular quantum slabs with $N=12\ldots 200$ as a 
function of the density parameter $r_s$.
}
\label{fig2}
\end{figure}

Next we focus on the rectangular QS and vary both $r_s$ and $N$.
The results are given in Fig.~\ref{fig2}
which, similarly to Fig.~\ref{fig1}, 
shows the relative total-energy errors of our functional (filled symbols)
and the TF (open symbols). The electron number
varies in the range $N=12\ldots 200$. As in the case of
a parabolic QD, the accuracy of our functionals is superior to that
of the TF, except at small $r_s$.

A significant feature in Fig.~\ref{fig2}
is the consistency of the accuracy of the present functional with $N$.
Instead, the validity of the TF approximation strongly depends 
on the electron number, which is due to the fact that at fixed $r_s$,
the relative amount of (classical) Hartree energy of the total energy
increases with $N$. However, even at large electron numbers our functional
is considerably more accurate than the TF approximation, on condition 
that the system is not too close to the noninteracting (small-$r_s$) 
regime. For example, at $N=120$ and $r_s\sim 3.3$, the relative
errors of the present functional and the TF approximation are $2\,\%$
and $10\,\%$, respectively. Hence, our functional is expected
to be a reliable tool for total-energy calculations in systems
that are computationally not reachable by, e.g., the LDA 
(see, e.g., Ref.~\onlinecite{afif}).
In fact, the $N=200$ case in Fig.~\ref{fig2} was close to the 
numerical limit of our LDA calculations.

We observed numerically that in 
QD's the difference between $T_{\rm TF}$ and $T_{\rm KS}$ 
is very small. As it is known, it actually goes to zero in the limit 
$r_s\rightarrow 0$ (Ref.~\onlinecite{Brack}). In QS's, on the other 
hand, $T_{\rm TF}$ may largely underestimate $T_{\rm KS}$. However, the relative
contribution of this underestimation to the total energy
strongly decreases as a function of $N$. It remains to be seen 
whether a fully self-consistent 
application of the presented functional may be carried out providing 
either accurate or at least better densities than the Thomas-Fermi ones.

\section{Conclusions}

In summary, we have derived an explicit 
density functional for the total energy of electrons
in two dimensions. The functional is numerically highly
efficient due to the freedom from orbitals and from
the calculation of the Hartree integral.
When applied to models of semiconductor quantum dots and slabs up
to hundreds of electrons, and up to strong electron-electron 
interactions, we have found a good overall agreement with respect to the 
local-density approximation, and a significant 
improvement over the Thomas-Fermi approximation.
Natural future developments of the present work
include the spin-dependent generalization and
the capability to deal with dimensional crossovers 
(such as from two to three dimensions, or from two to one dimension).

\begin{acknowledgments}
This work was supported by the EU's Sixth Framework
Programme through the Nanoquanta Network of
Excellence (No. NMP4-CT-2004-500198) and ETSF e-I3,
the Academy of Finland, and the Deutsche 
Forschungsgemeinschaft.
\end{acknowledgments}

\end{document}